\documentclass[12pt]{article}
\usepackage{epsfig}
\usepackage{latexsym}
\usepackage{amsmath}
\usepackage{amssymb}
\begin{document}

\begin{flushright}
JINR--E1--2006--97 \\
\end{flushright}

\vspace{8mm}
\begin{center}
\large \textbf{Search for gluinos with ATLAS at LHC}

\vspace{10mm}

\large V.A. Bednyakov$^{1}$, J.A. Budagov$^{1}$,
A.V. Gladyshev$^{2}$, D.I. Kazakov$^{2}$,\\
G.D. Khoriauli$^{1,3}$, D.I. Khubua$^{1,4}$

\vspace{10mm}
\normalsize

\textit{$^{1}$Dzhelepov Laboratory of Nuclear Problems,
Joint Institute for Nuclear Research, \\
141980, 6 Joliot-Curie, Dubna, Moscow Region, Russian Federation\\[2mm]
$^{2}$Bogoliubov Laboratory of Theoretical Physics,
Joint Institute for Nuclear Research, \\
141980, 6 Joliot-Curie, Dubna, Moscow Region, Russian Federation\\[2mm]
$^{3}$Tbilisi State University, Tbilisi, Georgia\\[2mm]
$^{4}$Institute of High Energy Physics and Informatics, \\
Tbilisi State University, Tbilisi, Georgia}

\end{center}

\vspace{10mm}

\begin{abstract}
Prospects for ATLAS observation of a SUSY-like signal from two
gluinos $pp \to \tilde g \tilde g$ are investigated within a certain
region of the mSUGRA parameter space, where the cross section of the
two gluinos production via gluon-gluon fusion, $gg \to \tilde g
\tilde g$, is estimated at a rather high level of 13 pb. The event
selection trigger  uses a very clear signature of the process (4
jets + 4 muons + up to 4 secondary vertices topology) when final
decay products of each gluino are $b$--anti-$b$ and muon--anti-muon
pairs and the lightest SUSY particle, the neutralino. Rather high
transverse missing energy carried away by two neutralinos is an
essential signature of the event and also allows the relevant
Standard Model background to be reduced significantly. The
generation and reconstruction processes are performed by means of
the ATLAS common software framework ATHENA.
\end{abstract}

\newpage

The minimal supersymmetric version of the Standard Model with
universal soft supersymmetry breaking terms induced by supergravity
(mSUGRA)~\cite{msugra} has a minimal set of free parameters:
$$
m_0, \ m_{1/2}, \ \mbox{sign}\,\mu, \ A_0, \ \tan\beta=v_2/v_1.
$$
where $m_0$  and $m_{1/2}$ are respectively common masses for scalar
and spinor superpartners at the unification scale, $\mu$  is the
Higgs mixing parameter, $A_0$ is the trilinear soft supersymmetry
breaking parameter and $\tan\beta$ is the ratio of vacuum
expectation values of two Higgs fields. The allowed values of these
parameters are limited by both theoretical and experimental
constraints as well as recent astrophysical data~\cite{wmap}.

In supersymmetric models with conserved $R$-parity, superpartners of
ordinary particles may be created only in pairs, which leads to the
existence of the stable lightest supersymmetric particle (LSP).
Usually it is the lightest neutralino -- neutral, massive and weakly
interactive particle. The mass spectrum of superpartners at some
scale in mSUGRA can be calculated by solving renormalization group
equations with parameters mentioned above being initial conditions
at the unification scale.

\begin{figure}[phtb]
\centerline{\psfig{file=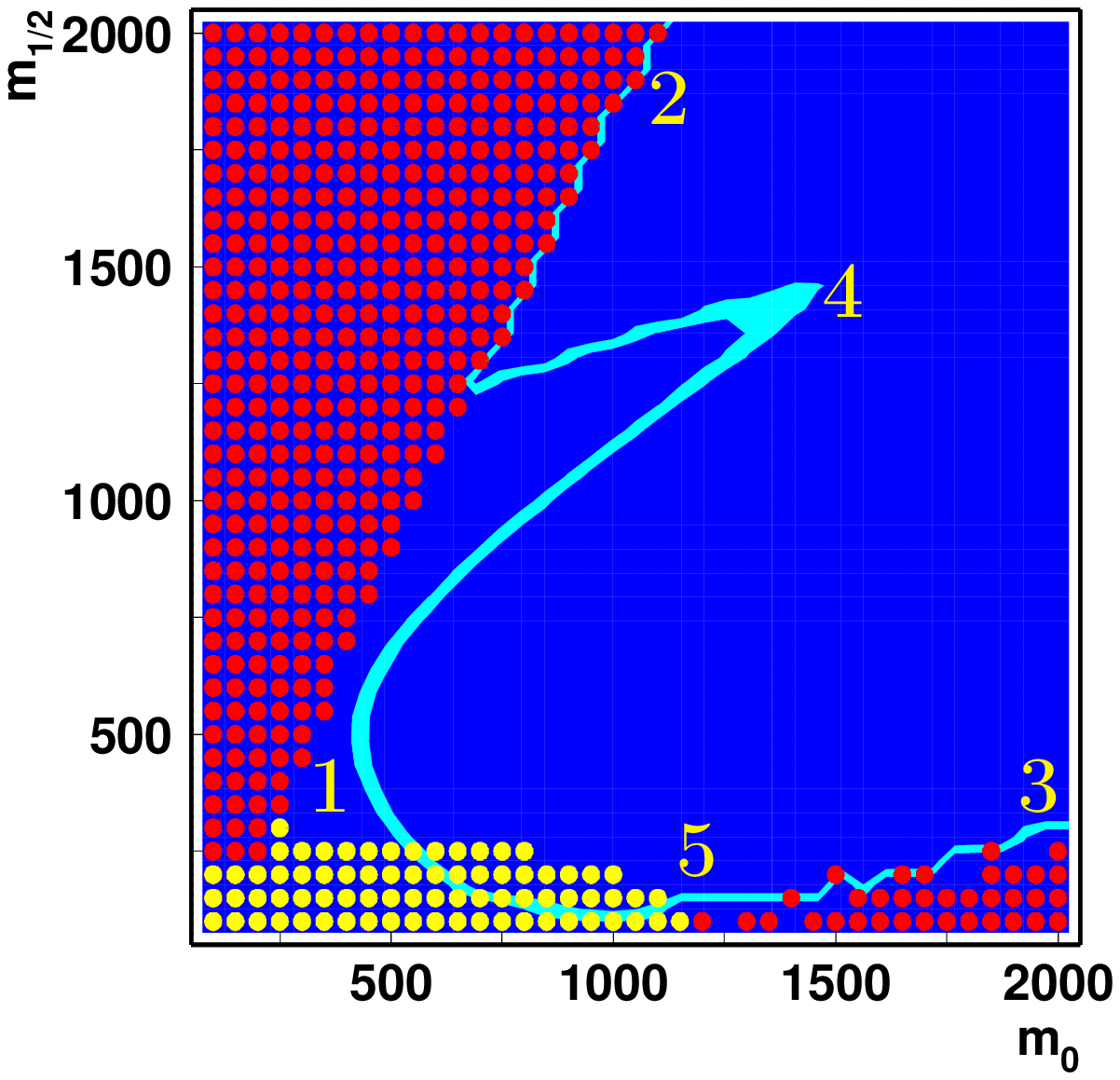,width=0.47\textwidth}
\psfig{file=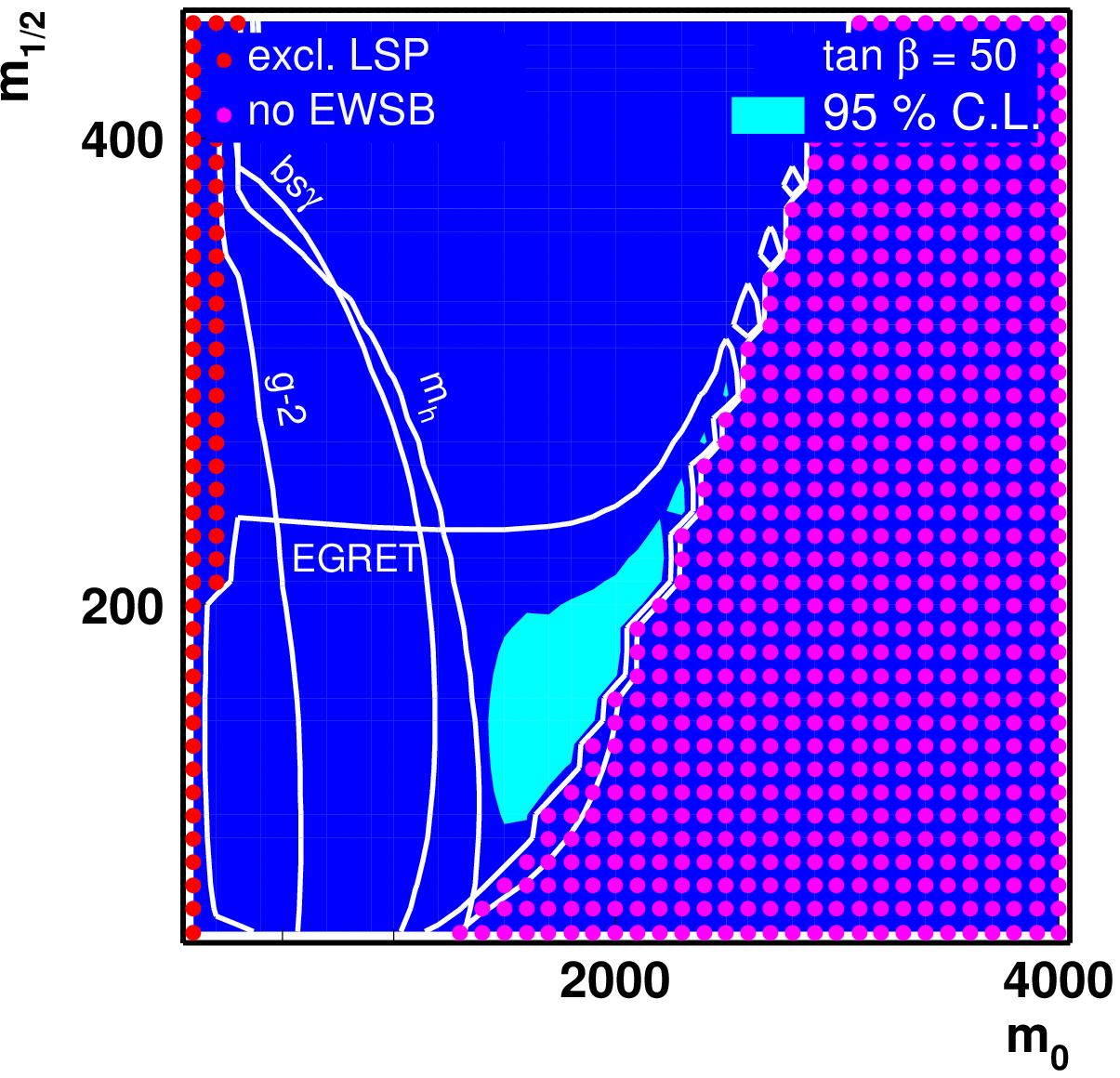,width=0.47\textwidth}}
\vspace*{8pt}
\caption{Left: MSSM parameter space. Light blue line corresponds to
the region consistent with WMAP data. One can distinguish between
different regions marked by digits. 5 -- is the EGRET region. Right:
The light shaded (blue) area is the enlarged region preferred by
EGRET data for $\tan\beta=50$, $\mu>0$ and $A_0=0$. The excluded
regions where the stau would be the LSP, or the electroweak symmetry
breaking fails, or the Higgs boson is too light are indicated by the
dots. \label{fig:fig1}}
\end{figure}

In the present analysis the region with large scalar masses $m_0$
and small fermion masses $m_{1/2}$ was studied. In this case LSP
properties are consistent with observed relic density, i.e. they can
serve as cold dark matter particles, and, on the other hand, can
naturally explain the excess of diffuse galactic gamma rays observed
by EGRET~\cite{egret}. The set of the parameters used in the
analysis (let us call it the EGRET point) is the following
(Fig.\ref{fig:fig1}):
$$
m_0= 1400 \ \mbox{GeV}, \ m_{1/2}=180 \ \mbox{GeV}, \
\mbox{sign}\,\mu=+1, \ A_0=0, \ \tan\beta=50.
$$
\begin{figure}[phtb]
\centerline{\psfig{file=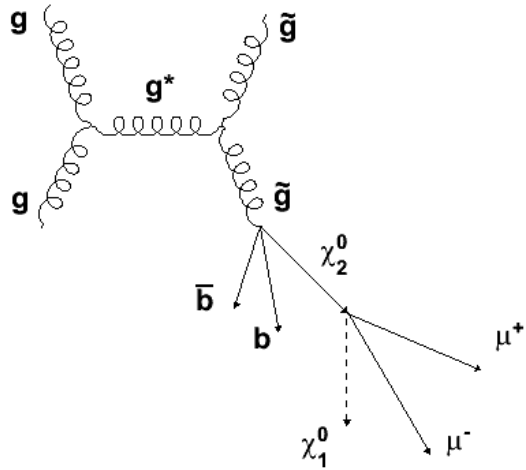,width=0.5\textwidth}}
\caption{Topology of the "half of the event" under study with 2
$b$-jets, 2 muons and 2 secondary vertices. \label{fig:fig2}}
\end{figure}
This region has not been intensively studied yet, though it is
within the reach of Tevatron and forthcoming LHC. It appears to be
very interesting phenomenologically, because of the mass splitting
between light gauginos and heavy squarks and sleptons. The
cross-sections for chargino and neutralino production in this case
are relatively large not being suppressed by masses and being
comparable with squark and gluino production. The latter being
enhanced by strong interactions remains suppressed by heaviness of
squarks. This means that in the EGRET region leptonic channels are
not suppressed and so might give  clear leptonic signature for
supersymmetry in the upcoming LHC experiments~\cite{we_ijmpa}.

The detection of a SUSY-like signal at LHC corresponding to the
EGRET point of mSUGRA would be a strong indication both for
supersymmetry and for solution of the dark matter problem. These
parameters (EGRET point) were used as input for the ISAJET
code~\cite{isajet} which calculated the superparticle spectrum.
Later the PYTHIA generator~\cite{pythia} used the spectrum for event
generation. It is worth mentioning that the whole generation process
was performed within the ATLAS software ATHENA~\cite{athena}.
Superparticle spectrum and cross-sections of different processes
depend on chosen parameters of mSUGRA. The common property of
SUSY-like processe in models with conserved $R$-parity is
undetectable LSP in the final state which are rather heavy and thus
take away quite high (and undetectable) transverse momentum. The
condition for choosing of the certain SUSY channel was not only the
large cross-section but also a peculiar signature which would tell
it from different Standard Model background events. In the EGRET
point the process of pair gluino production and their subsequent
decay appeared to be the most interesting:
\begin{equation}
\begin{array}{rrr}
gg \to \tilde g \tilde g & & \\
\downarrow & & \\
&\!\!\!\!\!\!\!\! \overline b + b + \tilde\chi^0_2 & \\
&  \downarrow \ & \\
& & \!\!\!\!\!\!\!\!\!\! \mu^- + \mu^+ + \tilde\chi^0_1 \\
\end{array}
\vspace{2mm}
\label{process1}
\end{equation}

Here we assume that both gluinos decay in the same way. So, in the
final state the gluino pair give 4 $b$-quarks ($b$-jets), 4 muons
and a pair of the lightest stable neutralinos $\tilde\chi^0_1$
giving the high missing transverse momentum. There are $B$-hadrons
in these jets, and, in general, the event could have 4 secondary
vertices, which allows to reduce the background even at the trigger
level (Fig.~\ref{fig:fig2}).

\begin{figure}[tp]
\centerline{\psfig{file=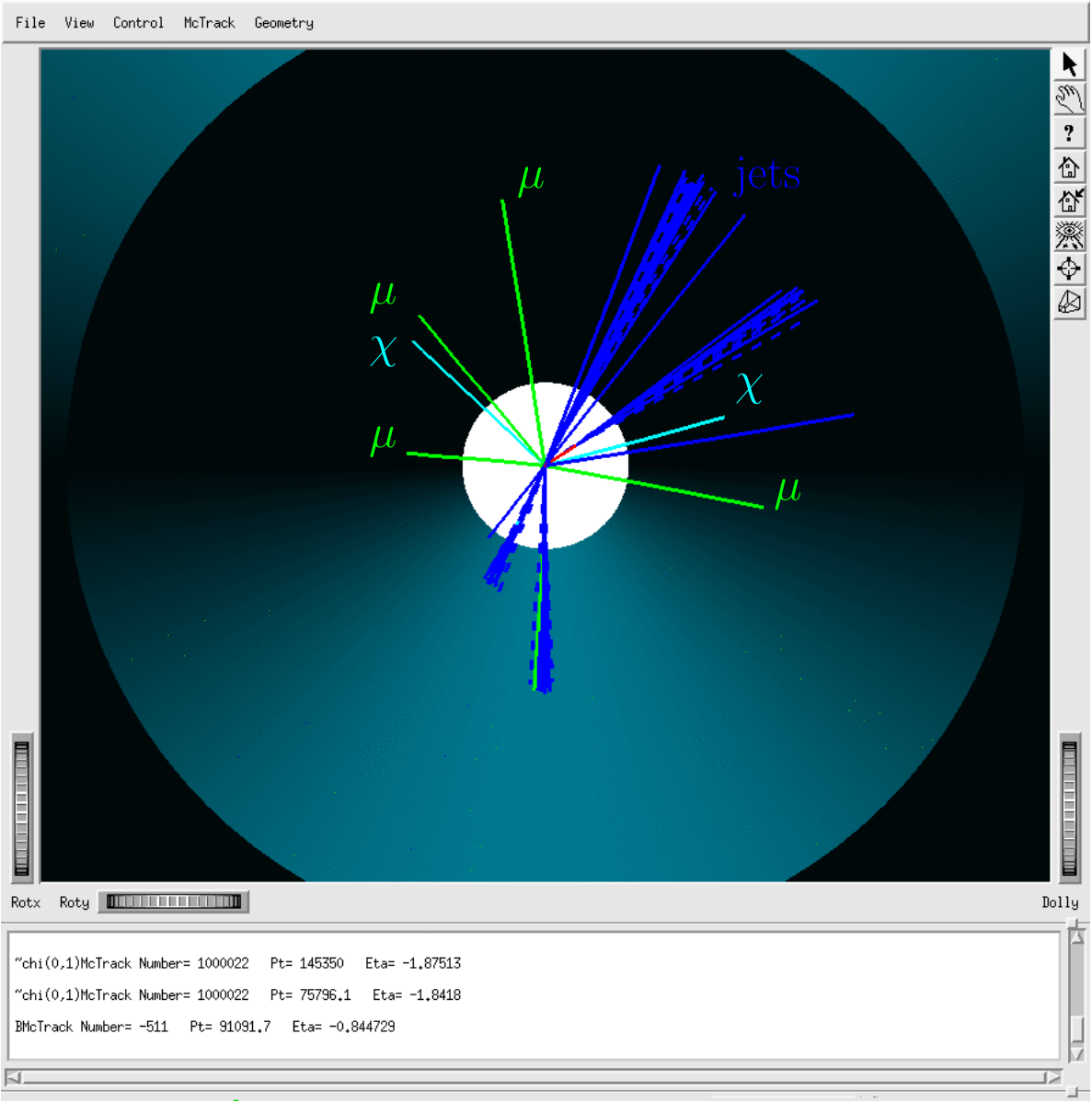,width=\textwidth}}
\caption{Generation of the process~\ref{process1} inside the
cylindrical (Layer 1) pixel detector in the $XY$-plane (transversely
to the beam). One can see 4 muon tracks (green lines), 2 tracks from
neutralino (light blue lines), 4 jets (dark blue lines) and one
long-living $B$-meson (red line). Visualisation has been performed
with the help of ATLAS GeoModel HitDisplay~\cite{geomodel}.
\label{fig:detector}}
\end{figure}

Fig.~\ref{fig:detector} shows the event inside the ATLAS
pixel detector in cylindrical first layer ($R \approx 4$ cm).
The cross-section of the hard $gg \to \tilde g \tilde g$ process at
$\sqrt{s}=14$~TeV is quite large. It varies between 5.6--14.2~pb
(PYTHIA) at the chosen values of SUSY parameters (EGRET point). The
cross-section depends on the parton distribution function (PDF) for
the proton and its dependence on $Q^2$.  We chose the
independent on $Q^2$ PDF model GRV94D (DIS)~\cite{pdf} and put
$Q^2_{min}=0.5$~GeV. In this case the cross section is $\approx$ 13
pb.

\begin{figure}[tbp]
\centerline{\psfig{file=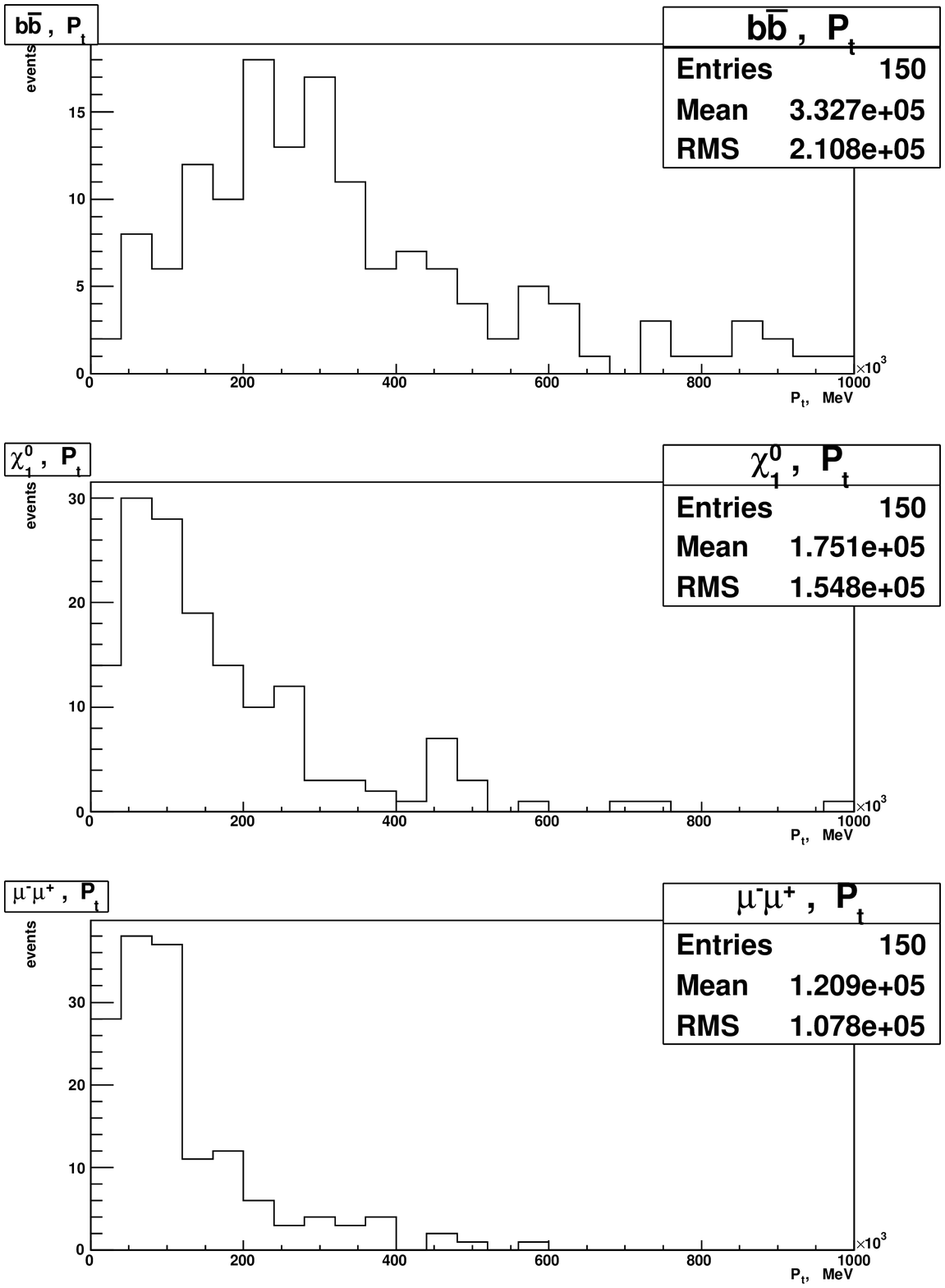,width=0.9\textwidth}}
\caption{Transverse momentum distributions for the products of one
of the gluinos decay. Top: the total $P_t$ for the $b\overline b$
pair; Middle: $P_t$ for the lightest stable neutralino
$\tilde\chi^0_1$ from the heavier neutralino decay; Bottom: total
$P_t$ for the muon pair. \label{fig:distributions}}
\end{figure}
\begin{figure}[t]
\centerline{\psfig{file=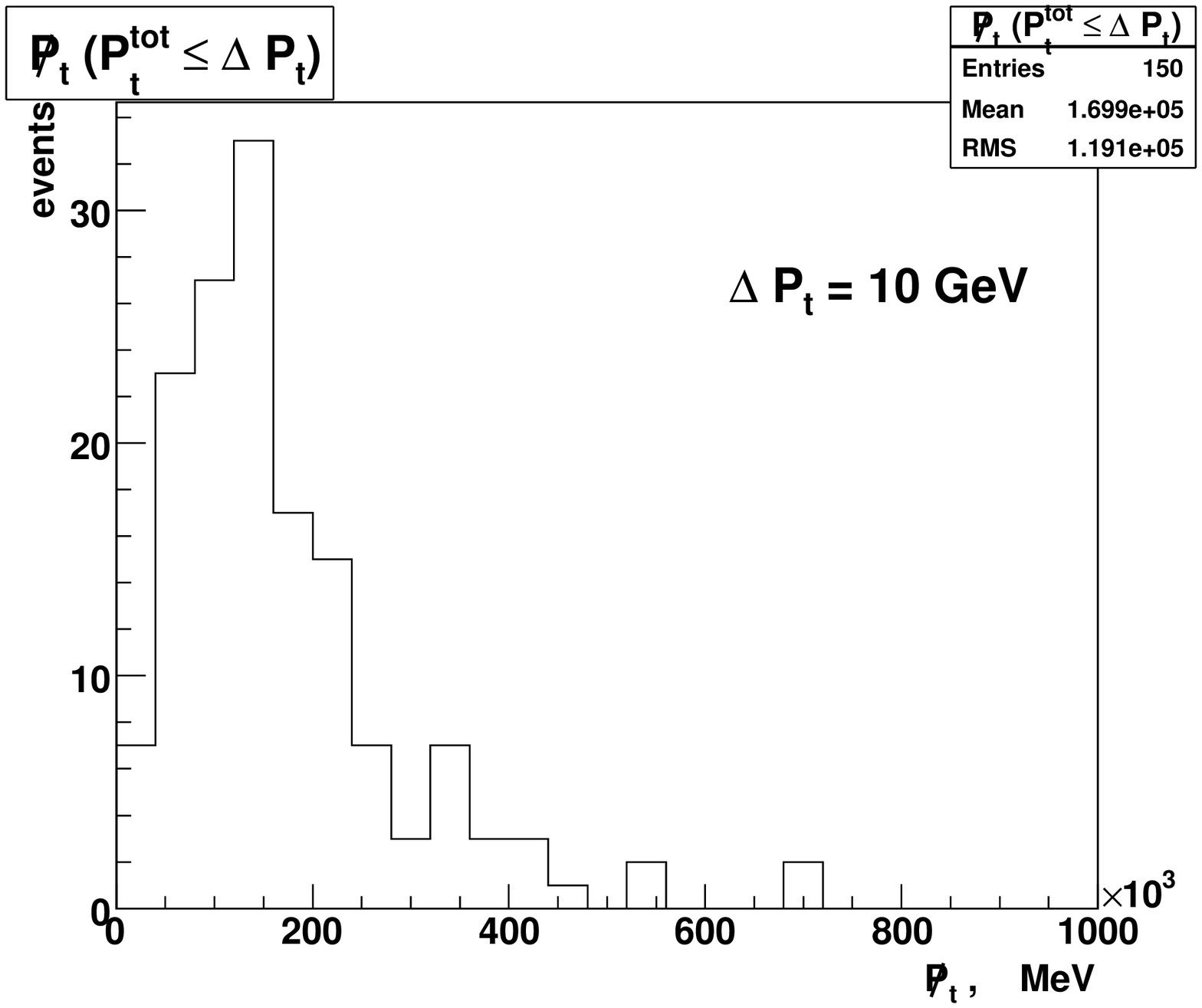,width=0.95\textwidth}}
\caption{Total missing transverse momentum $P_t$ of two neutralinos.
Event selection is made assuming that the total $P_t$ of gluino pair
is less than 10 GeV. Since in our study the initial state radiation
was not taken into account, all events satisfy this condition.
\label{fig:neutralinos}}
\end{figure}

The transverse momentum $P_t$ distribution for one gluino decay
products (two $b$-quarks forming 2 $b$-jets as well as 2 muons and
the lightest neutralino) is shown in Fig.~\ref{fig:distributions}.
One can see that neutralino takes away quite high transverse
momentum. It is worth noticing that in the real experiment one can
measure only the total missing (undetectable) energy of two
neutralinos $E_t \!\!\!\!\! \slash$. Fig.~\ref{fig:neutralinos}
shows the total transverse momentum of two neutralinos. Clearly,
that the careful reconstruction will allow to detect such a high
loss in the total measure transverse energy.

\begin{figure}[htb]
\centerline{\psfig{file=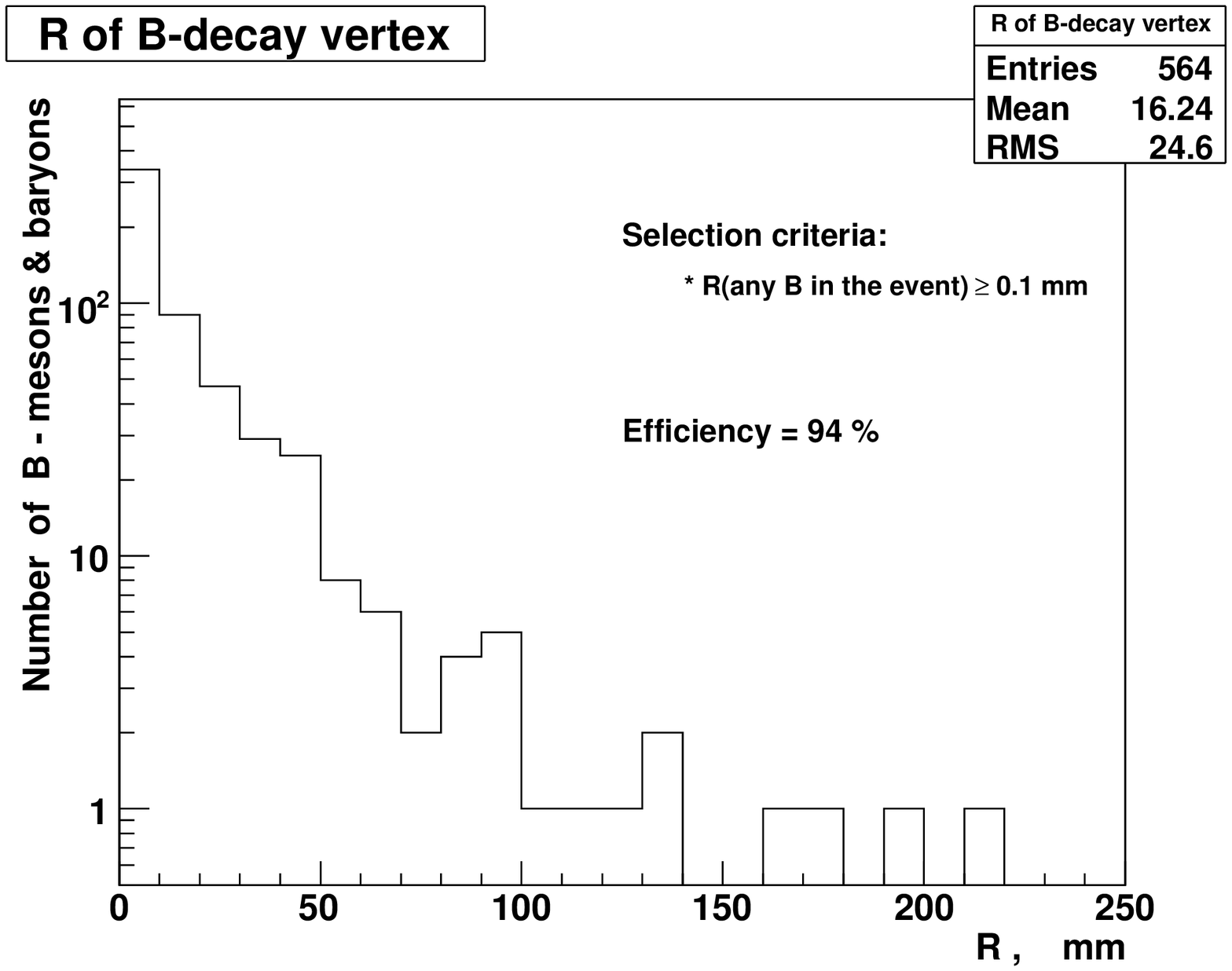,width=0.95\textwidth}}
\caption{The free path of $B$-hadrons created in the process
(\ref{process1}) before their decay. The criterion for
the event selection is that the free path is not less than
100 $\rm{\mu m}$ for each of four $B$-hadrons in the same event.
94\% of events satisfy this condition.
\label{fig:freepath}}
\end{figure}

The $b$-tagging of all $b$-jets and the careful reconstruction of
their energy are extremely important for the study of the considered
processes. $B$-hadrons with $b$-quarks inside live rather long time
and can move rather far away off the creation point (the cross point
of initial proton beams). As a result it allows to observe a
secondary vertex of $B$-hadron decay at a certain distance from the
primary beams collision initial vertex. This secondary vertex allows
to tag hadronic jets from $b$-quarks. Fig.~\ref{fig:freepath} shows
the distribution of the free path of $B$-hadrons provided all four
$B$-hadrons have free paths more than 100 $\rm{\mu m}$ simultaneously.
One can see that 94\% of events satisfy this condition.

All above presented distributions were performed for 150 events,
which is expected for ATLAS detector after a year of running with
total LHC luminosity $10^{34}~\rm{cm}^{-2}\rm{s}^{-1}$ for chosen
values of SUSY parameters, parton distribution functions and their
$Q^2$ dependence. Other PYTHIA parameters were set to their default
values. The process~(\ref{process1}) has the very peculiar signature
(4 $b$-jets, 4 muons and high transverse momentum), however there is
a mimic background process with only Standard Model particles
(top-quarks and gluons) with alike signature:
\begin{equation}
\begin{array}{rllll}
gg (q \overline q)\to t & \overline t \ \ \ t \ \ \ \overline t & & & \\
\; \raisebox{2pt}{$\Bigl\downarrow$} &
\!\!\! \begin{array}{l}
\downarrow \\
\overline b + W^-(\to \mu^- + \nu_\mu) \\
\end{array} & & \\
& \!\!\!\!\!\!\!\!  b + W^+(\to \mu^+ + \overline\nu_\mu) & & & \\
\end{array}
\vspace{2mm}
\label{process2}
\end{equation}

In this process~(\ref{process2}) we again have 4 $b$-quarks (4
$b$-jets), 4 muons and the missing transverse momentum, but from 2
neutrinos. Now this background process is under study, however we
can make some qualitative conclusions.
The cross-section of the processes
$gg(q\overline q) \to t \overline t t \overline t$
varies (depends on choice of $Q^2$-scale) in the range of
2.6 -- 9.4 (0.5 -- 1.5) fb (Acer MC-generator~\cite{acer}).
So, at the total LHC luminosity, 10$^{34}$ cm$^{-2}$s$^{-1}$,
the expected rate of
$gg,q\overline q \to t \overline t t \overline t
(W^+ b W^- \overline b W^+ b W^- \overline b)$ is up to 3500
events per year. We require that all of 4 $W$-bosons decay into
muon--neutrino pair ($BR(W^- \to \mu^- \nu_\mu \approx 0.1082)$)
to give the same signature as one for the SUSY process~\ref{process1}.
Then the SM background (process~\ref{process2}) rate is $< 0.5$
events per year. So, the preliminary estimation of the SM background
gives its negligible contribution in the total signal.
Moreover, due to the negligible (0 in our analysis) neutrino mass
the missing transverse momentum is smaller than in the
process~(\ref{process1}). That is there is a possibility to
distinguish between the background~(\ref{process2}) and SUSY-like
process~(\ref{process1}) by choosing events with the large missing
transverse momentum (larger than 50 GeV). Now we perform the full
analysis (event simulation and reconstruction) of these processes
using ATHENA.

To conclude, the prospects for ATLAS observation of a SUSY-like
signal from two gluinos are under investigation. The region of the
mSUGRA relevant to SUSY dark matter interpretation of the EGRET
observation is considered. The cross section of the two gluinos
production via gluon-gluon fusion, $gg \to \tilde g \tilde g$, is
rather high in the region. The clear signature of the process is 4
$b$-jets, 4 muons and rather large missing transverse energy (due to
2 lightest SUSY particles, the neutralinos). This large transverse
missing energy carried away by two neutralinos allows the relevant
Standard Model background to be reduced significantly. The
generation and reconstruction processes are performed within ATHENA
ATLAS software.

\vspace{5mm}

Financial support from the Russian Foundation for Basic Research
(grants \# 05--02--17603, \# 06--02--04003) and the grant of the
President of Russian Federation for support of leading scientific
schools (NSh--5362.2006.2) is kindly acknowledged.

\end{document}